\documentclass{article}

\usepackage{microtype}
\usepackage{graphicx}
\usepackage{subfigure}
\usepackage{booktabs} % for professional tables
\usepackage[ruled,vlined,linesnumbered]{algorithm2e}
\usepackage{amsmath}
\usepackage{amssymb}
\usepackage{caption}
\DontPrintSemicolon

\usepackage{hyperref}
 \usepackage{array}
\usepackage{algorithmic}

\SetKwInput{Input}{Inputs}
\SetKw{Break}{break}
\SetKwComment{tcp}{\scriptsize$\triangleright$~}{}
\newcommand{\clip}{\operatorname{clip}}
\newcommand{\depth}{\operatorname{depth}}
\newcommand{\mem}{\operatorname{mem}}
\newcommand{\NearestGraph}{\textsc{NearestGraph}}
\usepackage[accepted]{mlsys2025}

\mlsystitlerunning{LAPS: A Length-Aware-Prefill LLM Serving System}

\begin{document}

\twocolumn[
\mlsystitle{LAPS: A Length-Aware-Prefill LLM Serving System}

% It is OKAY to include author information, even for blind
% submissions: the style file will automatically remove it for you
% unless you've provided the [accepted] option to the mlsys2025
% package.

% List of affiliations: The first argument should be a (short)
% identifier you will use later to specify author affiliations
% Academic affiliations should list Department, University, City, Region, Country
% Industry affiliations should list Company, City, Region, Country

% You can specify symbols, otherwise they are numbered in order.
% Ideally, you should not use this facility. Affiliations will be numbered
% in order of appearance and this is the preferred way.
\mlsyssetsymbol{equal}{*}

\begin{mlsysauthorlist}
\mlsysauthor{Jianshu She}{mbzuai}
\mlsysauthor{Zonghang Li}{mbzuai}
\mlsysauthor{Hongchao Du}{mbzuai}
\mlsysauthor{Shangyu Wu}{mbzuai}
\mlsysauthor{Wenhao Zheng}{unc}
\mlsysauthor{Eric Xing}{mbzuai}
\mlsysauthor{Zhengzhong Liu}{mbzuai}
\mlsysauthor{Huaxiu Yao}{unc}
\mlsysauthor{Jason Xue}{mbzuai}
\mlsysauthor{Qirong Ho}{mbzuai}
\end{mlsysauthorlist}

\mlsysaffiliation{mbzuai}{Mohamed bin Zayed University of Artificial Intelligence}
\mlsysaffiliation{unc}{The University of North Carolina at Chapel Hill}

\mlsyscorrespondingauthor{Jianshu She}{jianshu.she@mbzuai.ac.ae}
\mlsyscorrespondingauthor{Qirong Ho}{Qirong.Ho@mbzuai.ac.ae}

% You may provide any keywords that you
% find helpful for describing your paper; these are used to populate
% the "keywords" metadata in the PDF but will not be shown in the document
\mlsyskeywords{Machine Learning, MLSys}

\vskip 0.3in

\begin{abstract}

LAPS identifies and disaggregates requests with different prompt lengths in LLM serving to reduce TTFT latency. While recent systems have decoupled the prefill and decode stages to improve throughput, they still rely on unified scheduling policies that fail to adapt to heterogeneous workload characteristics. We observe that prompt-length variations lead to distinct performance bottlenecks, motivating an adaptive scheduling strategy. LAPS disaggregates multi-turn long-prefill requests from short-prefill ones and introduces a length-aware smart batching mechanism for short-prefill workloads. It adopts a dual-queue design that supports temporal disaggregation on a single prefill instance or spatial disaggregation across multiple instances. For short-prefill batches, a batch waiting window and CUDA Graph-based clustering mitigate interference from heterogeneous computation, reducing batching delay and lowering average latency. In real multi-turn workloads, LAPS reduces prefill latency by over 30\% compared to vanilla SGLang under prefill–decode disaggregation, and further decreases SLO violations by 28\% in multi-instance deployments with vanilla data-parallel configuration. Compared to the SGLang router with load balancing, it further lowers SLO violations by 12\% in multi-GPU settings. Under high concurrency and mixed-request scenarios, LAPS improves request throughput by 35\% serving Qwen2.5-32B model for prefill instance, demonstrating its effectiveness in optimizing heterogeneous LLM serving workloads.

\end{abstract}
]

% this must go after the closing bracket ] following \twocolumn[ ...

% This command actually creates the footnote in the first column
% listing the affiliations and the copyright notice.
% The command takes one argument, which is text to display at the start of the footnote.
% The \mlsysEqualContribution command is standard text for equal contribution.
% Remove it (just {}) if you do not need this facility.

%\printAffiliationsAndNotice{}  % leave blank if no need to mention equal contribution
\printAffiliationsAndNotice{\mlsysEqualContribution} % otherwise use the standard text.

\section{Introduction}

Modern LLM serving stacks (e.g., vLLM \cite{kwon2023efficientmemorymanagementlarge}, SGLang \cite{zheng2024sglangefficientexecutionstructured}) combine prefill-decoding (PD) disaggregation \cite{zhong2024distservedisaggregatingprefilldecoding} with continuous
batching to meet low-latency, high-concurrency service-level objectives (SLOs). The Prefill stage (first-token computation) is largely compute-bound, while the decoding stage (auto-regressive generation) is memory-bound. PD disaggregation decouples these two phases across separate instances to avoid cross-phase contention. On the prefill side,
concurrent requests are batched to raise GPU utilization. However, we show that this separation and batching are insufficient: even with PD, interference still persists within the prefill stage when long, compute-bound prefills are mixed with short, memory-bound prefills/re-prefills.

\begin{figure}[h]
    \centering
    \includegraphics[width=\linewidth]{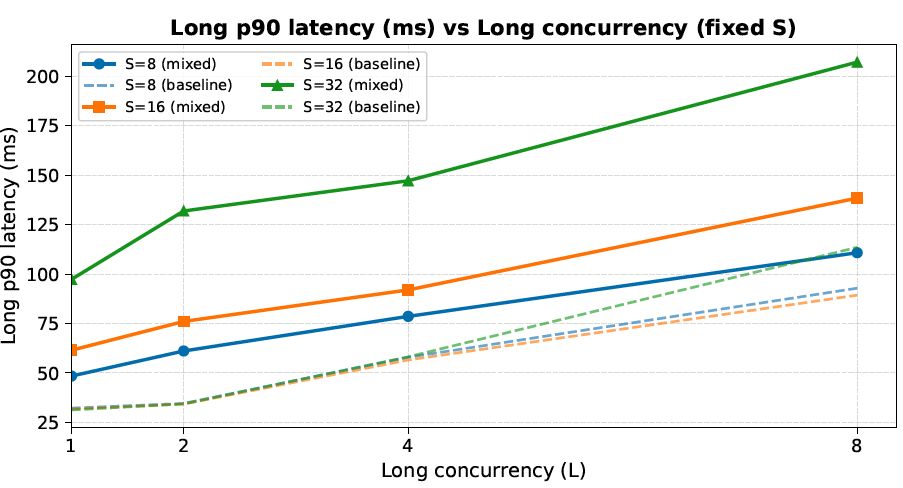}
    \captionsetup{aboveskip=2pt, belowskip=-5cm}
    \caption{P90 TTFT of long-prefill requests under varying concurrency levels for long and short requests. The long-prefill requests have more than 1K tokens, while the short ones have fewer than 64 tokens. We concurrently run them on a single H200 GPU and serve by Qwen2.5-32B \cite{qwen2025qwen25technicalreport}. The dashed lines indicate the latency when only long-prefill requests are served.
    % We continuously issue long-prefill requests (exceeding 1K tokens) and short ones (fewer than 64 tokens) on a single H200 GPU, serving Qwen2.5-32B model \cite{qwen2025qwen25technicalreport}. We measure the P90 latency of TTFT under varying concurrency levels for long and short requests, where the dashed lines indicate the TTFT P90 latency when only long requests are present.
    }
    \label{fig:interfere}
\end{figure}

\iffalse
\begin{figure}
    \centering
    \includegraphics[width=0.9\linewidth]{Image/workload.png}
    \caption{The horizontal axis represents input (prefill) length, and the vertical axis represents output (decode) length. We categorize LLM serving workloads into four quadrants: Short Prefill, Short Decode (SPSD), Short Prefill, Long Decode (SPLD), Long Prefill, Short Decode(LPSD), and Long Prefill, Long Decode (LPLD).}
    \label{fig:workload}
\end{figure}
\fi

\begin{figure*}
    \centering
    \includegraphics[width=0.9\linewidth]{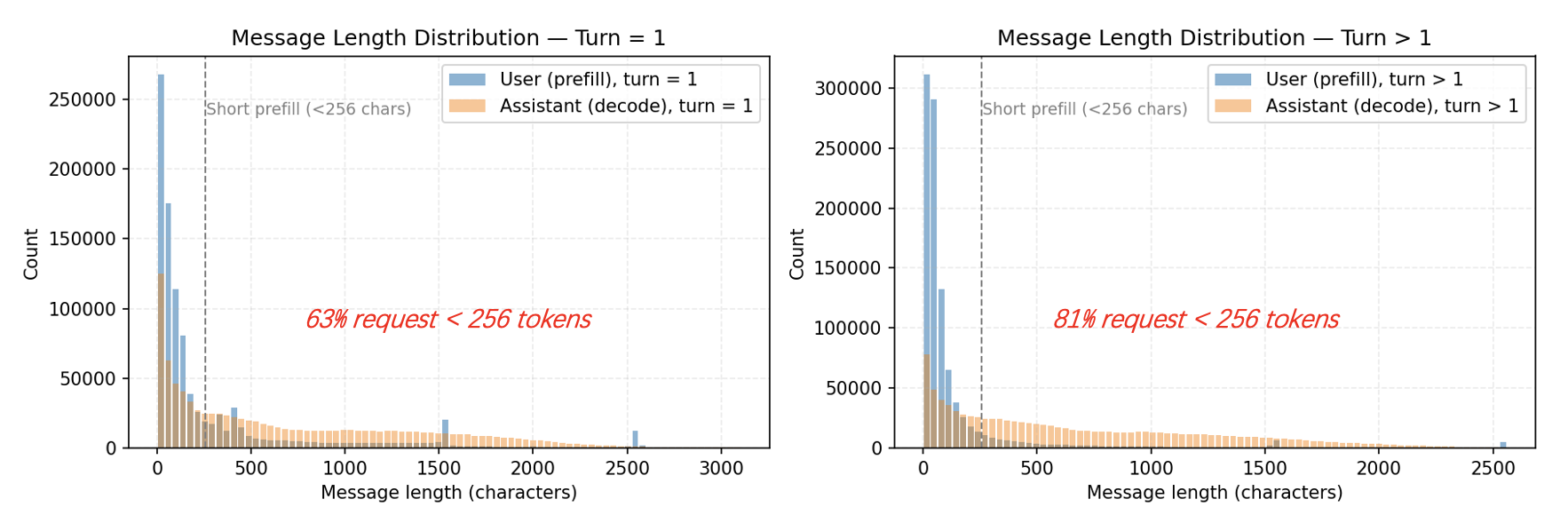}
    \vspace{-5pt}
    \caption{The token length distribution of multi-turn dialogues in the real LMsys-Chat-1M dataset. The left plot shows the prompt length in the first turn (including the system prompt by default), where approximately 63\% of requests contain fewer than 256 tokens. In subsequent turns, the proportion of prompts shorter than 256 tokens increases to an average of 81\%.}
    \label{fig:dataset distribution}
\end{figure*}

Re‑prefill denotes the repeated prefill in multi‑turn sessions where the model extends an existing context by
combining new tokens with cached KV states. It is common in chatbots~\cite{dam2024completesurveyllmbasedai}, tool-using agents~\cite{wölflein2025llmagentsmakingagent}, RAG, speculative
decoding~\cite{leviathan2023fastinferencetransformersspeculative}, and token routing~\cite{she2025tokenlevelroutinginference}, and is typically memory-bound (dominated by KV-cache reads/writes rather than
large GEMMs). Figure \ref{fig:dataset distribution} illustrates the token length distribution of a real-world trace, LMsys-Chat-1M \cite{zheng2024lmsyschat1mlargescalerealworldllm}, which is collected from real human-AI conversations. We observe that most prompts are short (\textless 256 tokens), while long-context requests (\textgreater 1K tokens) are relatively rare. This indicates
that production workloads are dominated by short prefills/re-prefills, which are memory-bound. Mixing them
with long, compute-bound prefills in unified batches will also induce compute-memory interference: short
prefills/re-prefills wait behind long GEMMs and time-to-first-token (TTFT) spikes, and long prefills lose effective
FLOPs due to heavy KV traffic from short jobs. 
Figure \ref{fig:interfere} confirms this issue: mixing long and short prefills does significantly increase long-prefill latency, and this contention becomes more severe as concurrency rises. 

Although prior systems have recognized the resource contention between compute-bound and memory-bound workloads, they focus on coordinating prefill and decode: (1) decode‑priority schedulers\cite{kwon2023efficientmemorymanagementlarge} prioritize the decode phase to minimize per-token latency; (2) prefill-first schedulers prioritize the prefill phase and use continuous batching~\cite{yu2022orca} to improve throughput; (3) stall-free chunked prefill \cite{agrawal2023sarathiefficientllminference} splits long prefill into chunks and interleaves them with decode, so long prefills don't stall ongoing decodes; and (4) PD disaggregation \cite{zhong2024distservedisaggregatingprefilldecoding,jin2024pdserveservingdisaggregatedlarge,hu2024inferenceinterferencedisaggregatellm,strati2024dejavukvcachestreamingfast}, which
dominates in modern serving systems, runs prefill and decode on separate instances. These advances alleviate
cross‑phase contention but implicitly assume all prefills are compute‑bound long sequences, overlooking that
short (e.g., \textless 256 tokens), memory‑bound prefills/re‑prefills could dominate real‑world multi-turn serving workloads.

%Existing systems only partially address this problem. Sarathi-Serve~\cite{agrawal2023sarathiefficientllminference} introduces \emph{chunked prefill} to interleave long and short requests, while DistServe~\cite{zhong2024distservedisaggregatingprefilldecoding} disaggregates prefill and decode across GPUs to reduce cross-phase interference. SplitServe~\cite{patel2024splitwiseefficientgenerativellm} similarly disaggregates the prefill and decode stages to improve system throughput under high-concurrency workloads 
%这里有一些算法层面上的长度感知. However, neither approach considers intra-prefill heterogeneity in multi-turn inference---specifically, the conflicting resource characteristics between short-prefill and long-prefill workloads within the same phase. As a result, mixed workloads continue to exhibit suboptimal throughput and inflated tail latency (TTFT/TPOT). Although some recent works attempt to make the serving system prompt-length aware by prioritizing short requests, there is still a lack of systematic investigation into the interference between long and short prefills in multi-turn workloads, as well as optimized scheduling strategies for high-frequency, GPU-unfriendly short-prefill requests.

To address these challenges, we propose LAPS, a length-aware LLM serving system that explicitly disaggregates and optimizes long-prefill and short-prefill workloads within the prefill stage. LAPS maintains two separate prefill pools at runtime and performs batch disaggregation to isolate long and short prefill requests, completely eliminating their mutual interference.
For short-prefill requests, LAPS introduces a dynamic waiting window in the scheduler and bucketizes requests by input length, making them more uniform and enabling larger batch sizes. This reduces batch-launch overhead and, combined with CUDA Graph-based execution, further accelerates processing and improves throughput.
For SLO-serving scenarios, we design an SLO-aware scheduler that balances the trade-off between the waiting window and throughput.

In multi-instance deployments, the scheduler dynamically adjusts each instance’s workload type based on real-time load, achieving load balancing across spatially disaggregated prefill instances.
This adaptive strategy resembles resource allocation problems in deep learning systems, like in \cite{qiao2021pollux}.

In addition, LLM serving systems typically support three modes: mix and PD temporal/spatial disaggregation. Beyond these, our LAPS introduces a fourth mode:
\begin{itemize}
    \item \textit{Mix:} Decode requests are inserted into prefill batches (without disaggregation);
    \item \textit{PD temporal disaggregation:} Prefill and decode batches run sequentially on the same instance;
    \item \textit{PD spatial disaggregation:} Prefill and decode batches run on separate instances.
    % \item \textit{Prefill temporal/spatial disaggregation:} Enabling prefill batches themselves to be temporally or spatially separated.
    \item \textit{Prefill batch temporal/spatial disaggregation (ours):} Enabling disaggregation within the prefill stage (rather than between prefill and decode), separating long- and short-prefill batches temporally on a single instance or spatially across instances.
\end{itemize}

It is worth noting that our LAPS remains compatible with existing PD disaggregation architectures. Overall, our main contributions are summarized as follows:
\begin{enumerate}
\item \textbf{Empirical characterization.} We analyze intra-prefill interference between long and short requests in multi-turn workloads, exposing compute-memory contention caused by current batching strategies.
\item \textbf{Length-aware disaggregation.} We design a request-level temporal/spatial prefill disaggregation architecture that isolates long and short prefills to eliminate interference.
\item \textbf{Adaptive scheduling.} We introduce a dynamic bucket-based batching policy with a waiting window and load balancing across prefill instances, improving throughput and reducing SLO violations.
\end{enumerate}

\section{Background and Motivations}
%multi-turn prefill的分析放在这里，interfere的实验设计和结果也放在这里
In this section, we model the token-length conditions under which prefill and re-prefill become memory-bound, how compute-bound long prefills and memory-bound short (re-)prefills interfere with each other, and what causes this long/short mixing.

\subsection{Compute-Memory Boundary for (Re-)Prefills}

The prefill and re-prefill phases have different latency behaviors. In re-prefill, the model processes new prompt tokens while also attending to historical tokens. This increases both computing and memory overhead, and shifts the token-length boundary (critical point) $L_m$ at which (re-)prefills transition from compute-bound to memory-bound. We now formulate a unified latency model to find the token-length boundaries $L_m^{\text{prefill}}$ and $L_m^{\text{re-prefill}}$ for prefill and re-prefill phases, respectively. We will use this model to show that {\bf both prefill and re-prefill remain memory-bound for shorter fill lengths}.

Let $L$ be the number of new tokens in this turn, $H$ be the number of historical tokens, and
$T(L,H) = T_{\text{comp}}(L,H) + T_{\text{mem}}(L,H)$ be the total latency.
The compute term reflects incremental causal attention and FFN:
\[
T_{\text{comp}}(L,H) \approx \alpha L (L + 2H) + \beta L,
\]
where $\alpha,\beta$ are per-token costs for attention and FFN compute, respectively.
The memory term models the time for KV read/write I/O:
\[
T_{\text{mem}}(L,H) \approx \gamma_w L + \gamma_r H,
\]
where $\gamma_w$ and $\gamma_r$ are per-token KV write/read times.

\paragraph{Prefills.}
In the first-turn prefill, there is no history ($H=0$), so
$T_{\text{comp}}(L,0) \approx \alpha L^2 + \beta L$
and $T_{\text{mem}}(L,0) \approx \gamma_w L$.
The boundary can be obtained by equating these two contributions, yielding:
\[
L_m^{\text{prefill}} = \max\!\left(0, \frac{\gamma_w - \beta}{\alpha}\right).
\]
If $\gamma_w \le \beta$, prefills are always compute-bound; otherwise, memory access dominates for small $L < L_m^{\text{prefill}}$.

\paragraph{Re-prefills.}
Similarly, for re-prefills with $H > 0$,
\[
T_{\text{comp}}(L,H) \approx \alpha L^2 + (2\alpha H + \beta)L,\quad
\]
\[
T_{\text{mem}}(L,H) \approx \gamma_w L + \gamma_r H,
\]
so the boundary is given by:
\begin{equation}
\nonumber
\resizebox{0.92\columnwidth}{!}{$
L_m^{\text{re-prefill}}=\max\!\left(0,\,
\frac{-(2\alpha H + \beta - \gamma_w)
+ \sqrt{(2\alpha H + \beta - \gamma_w)^2 + 4\alpha\gamma_r H}}{2\alpha}\right)
$}.
\end{equation}

For any fixed $H>0$, re-prefill is memory-bound for small $L<L_m^{\text{re-prefill}}$ because when $L\rightarrow 0$,
$T_{\text{mem}}(L,H)\rightarrow\gamma_r H>0$ while $T_{\text{comp}}(L,H)\rightarrow 0$.
As $H$ increases, the $L_m^{\text{re-prefill}}$ boundary grows until a saturation point: $L_m^{\text{re-prefill}} \to \frac{\gamma_r}{2\alpha}$ for large $H \gg |\beta - \gamma_w|/(2\alpha)$.
Thus, with long histories, re-prefills remain memory-bound up to a constant number of new tokens,
after which the $2\alpha H L$ and $\alpha L^2$ terms render the phase compute-bound.

\paragraph{Fitting at runtime.}
Compute and memory latency can be modeled as quadratic and linear functions of $(L,H)$, respectively.
We collect runtime samples $(T_{\text{comp}}, T_{\text{mem}}, L, H)$ to fit these two curves and
obtain $\alpha,\beta,\gamma_w,\gamma_r$, and then calculate the boundaries
$L_m^{\text{prefill}}$ and $L_m^{\text{re-prefill}}$.

\paragraph{Roofline model.}
We also use the arithmetic intensity and roofline model to characterize the transition between memory- and compute-bound workloads in the prefill stage.
The arithmetic intensity of prefill computation increases approximately linearly 
with the prompt length $L$, since longer sequences proportionally increase the ratio of arithmetic operations to memory access.
The compute-memory boundary occurs when the arithmetic intensity $AI(L)$ 
reaches the hardware roofline slope $AI^* = P_{\text{peak}} / B_{\text{mem}}$, 
where $P_{\text{peak}}$ and $B_{\text{mem}}$ denote the peak compute throughput 
and sustained memory bandwidth of the GPU.

Empirical profiling across advanced hardware generations (A100, H100, and H200) and LLMs ranging from 7B to 32B parameters shows that this transition typically occurs between 150 and 512 tokens \cite{yuan2024llminferenceunveiledsurvey,zhong2024distservedisaggregatingprefilldecoding}, depending on model architecture, kernel implementation, and batch configuration. 
Prefills shorter than this range tend to be memory-bound and limited by KV-cache I/O, while longer ones are dominated by GEMM throughput.
% roof line model

\subsection{Exploring Interference Between Long-Short Prefills/Re-prefills}

\begin{table*}[h!]
    \centering
\begin{tabular}{lll}
\toprule
\textbf{Category} & \textbf{Prefill / Decode Features} & \textbf{Typical Tasks} \\
\midrule
SPSD & Short prompt and short output &
Token-level routing, classification, short chat replies, tool selection \\
SPLD & Short prompt, long output &
Code generation, creative writing, long summarization, Chain-of-Thought \\
LPSD & Long context, short output &
Document QA, retrieval-based answering, chat reply, sentiment classification \\
LPLD & Long context and long output &
Long-form summarization, report writing, autonomous agents \\
\bottomrule
\end{tabular}
    \caption{Task classification by prefill and decode characteristics.
SPSD: short-prefill, short-decode;
SPLD: short-prefill, long-decode;
LPSD: long-prefill, short-decode;
LPLD: long-prefill, long-decode.}
    \label{tab:task classification}
\end{table*}

We model the interference between long-prefill and short-prefill (or re-prefill) requests
using a standard M/G/1 (Markovian/General/1-server queue) first-come, first-served (FCFS) queuing model \cite{meini1998solving}.
Unlike prior PD scheduling analyses that focused on prefill-decode contention,
we examine the \textit{intra-prefill} interference caused by mixing
compute-bound and memory-bound jobs within continuous batching.

In this model, each request passes through two service stations:
a compute station (for GEMM-dominated operations) and a memory station
(for KV-cache I/O).  
Short-prefill and re-prefill jobs are typically memory-bound, 
while long-prefill jobs are compute-bound.
Let the aggregate arrival rate be~$\lambda$ and utilization~$\rho < 1$.
Denote by~$p$ the fraction of short jobs in the workload.
The service time at the memory station is
$S_m = \gamma_w L + \gamma_r H$,
and at the compute station
$S_c = \alpha L^2 + (2\alpha H + \beta)L$.

\begin{figure}
    \centering
    \includegraphics[width=\linewidth]{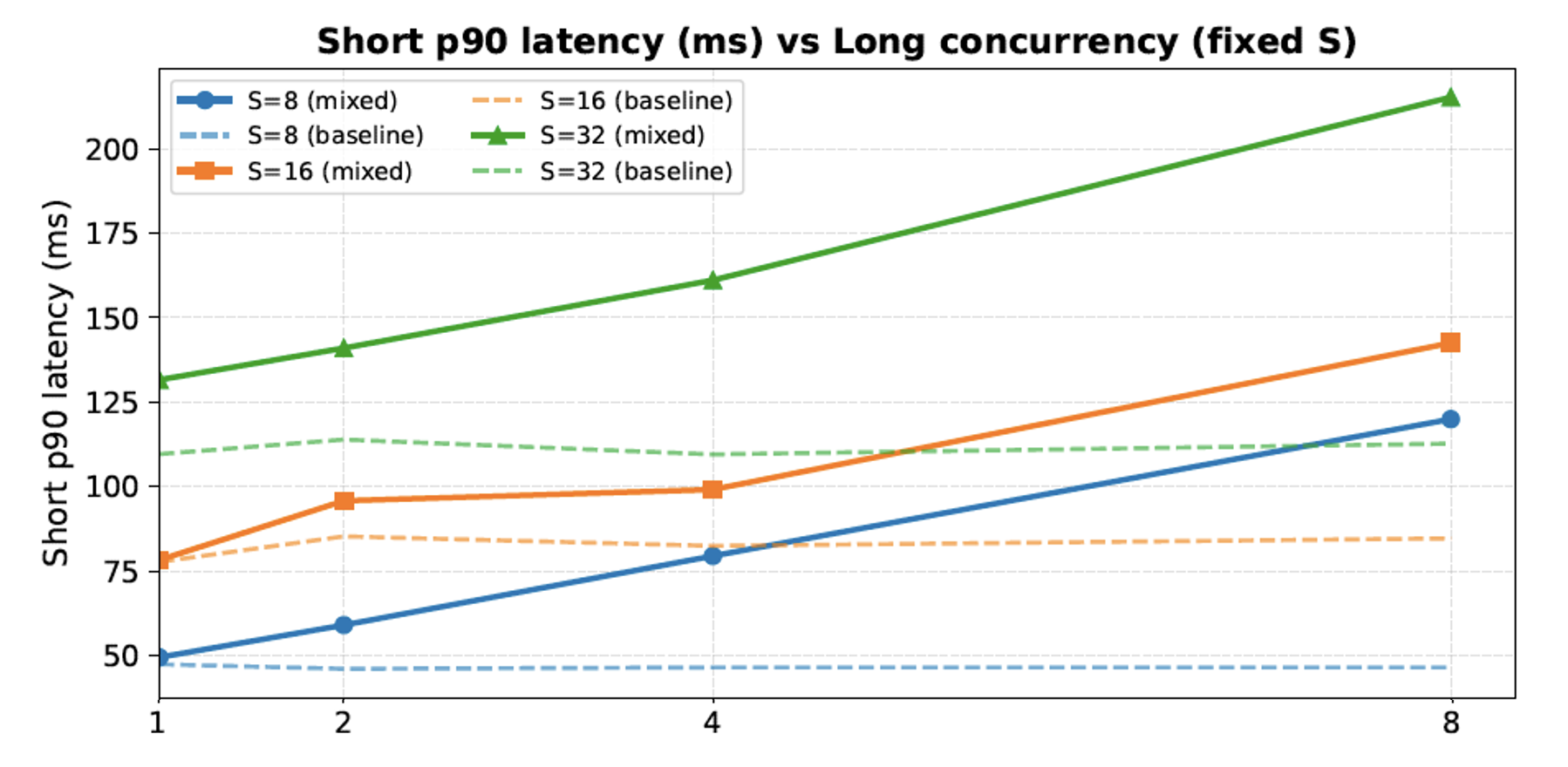}
    \caption{
    P90 TTFT of short-prefill requests under varying concurrency levels for long and short requests. The dashed lines indicate the latency when only short-prefill requests are served. Other setups are the same as those in Figure \ref{fig:interfere}.}
    % We continuously issue long prefill requests (exceeding 1K tokens) and short requests (fewer than 64 tokens) on a single H200 GPU, serving Qwen2.5-32B model. We measure the short request latency P90 latency of TTFT under varying concurrency levels for long and short requests, where the dashed lines indicate the TTFT P90 latency when only short requests are present.}
    \label{fig:interfere_1}
\end{figure}

Using the Pollaczek-Khinchine (P-K) result \cite{neuts1986generalizations} for M/G/1 queues, the mean waiting time is
\[
W = \frac{\lambda\,\mathbb{E}[S^2]}{2(1-\rho)}.
\]
When jobs of different lengths are batched together, the variance in service times inflates waiting time for all requests,
introducing a \textbf{head-of-line (HoL) blocking penalty}:
\[
\Delta W_{\text{HoL}} =
\frac{\lambda p(1-p)}{2(1-\rho)}(S_\ell - S_s)^2 > 0.
\]
This term grows with higher concurrency and service heterogeneity,
explaining the observed latency increase in mixed long/short-prefill workloads
(shown in Figures~\ref{fig:interfere} and~\ref{fig:interfere_1}).

Furthermore, long prefills hurt short (re-)prefills more.
Every class sees the same queuing term~$W$,
so normalized latency is $R_i / S_i = 1 + W/S_i$.
Given $S_s < S_\ell$, the relative increase is larger for short jobs
because $W/S_s > W/S_\ell$.
This convoy effect explains why short-prefill latency grows faster as long-prefill concurrency increases, which is a clear symptom of \textbf{bandwidth contention}.

\subsection{Uncovering the Sources of Long-Short Prefill Mixing}

General-purpose LLM services must handle a wide spectrum of tasks.
As shown in Table~\ref{tab:task classification}, daily chat and creative ideation
are typical short-prompt tasks, while speculative decoding and token routing
produce high-frequency, short re-prefills \cite{chatterji2025people}.
In contrast, long-document QA and autonomous agent workflows correspond to
long-context prefills.
In practice, these streams interleave over time, leading to
long-short mixing within prefill workloads.

Most existing systems schedule requests in an FCFS fashion, packing them into unified batches.
Many deploy multi-queue variants:
continuous or rolling batching (e.g., vLLM \cite{kwon2023efficientmemorymanagementlarge}, TGI \cite{huggingface_tgi_docs_2024}) treats prefill and decode
as distinct phases, applying FCFS-style admission under token/KV limits
and optional priorities or aging policies.
With chunked prefill (e.g., Sarathi-Serve \cite{agrawal2023sarathiefficientllminference}), vLLM prioritizes decode
and may co-batch prefills with decode.
In-flight batching (e.g., TensorRT-LLM \cite{tensorrt_llm}) runs distinct context and generation engines,
each maintaining its own ready queue and often prioritizing generation.
Moreover, many LLM gateways expose service tiers that prioritize
certain request classes while enforcing token-based budgets
using fixed or sliding windows (e.g., OpenAI \cite{openai_pricing}, Anthropic \cite{anthropic_pricing}, Envoy \cite{envoy_gateway}, Kong \cite{kong_gateway}, APISIX \cite{apisix_gateway}, and Cloudflare \cite{cloudflare_api}).

These systems indeed adopt multi-queue designs,
but most queues are phase-oriented (prefill vs.\ decode) or SLA-based.
Consequently, long and short (re-)prefills still end up co-batched
even under multi-queue scheduling.
Long-prefill requests have longer residence times,
and schedulers backfill every few milliseconds,
so newly arriving short (re-)prefills are co-admitted into the same batch.
Larger admission windows or micro-batches further raise the odds of co-admission,
while speculative decoding and token routing inject
frequent short jobs alongside ongoing long prefills.

The most related line of work is \emph{length bucketing}, which groups requests by predicted sequence length into size-homogeneous buckets to reduce padding and improve throughput (e.g., Multi-Bin Batching \cite{guldogan2024multi}, BucketServe \cite{zheng2025bucketserve}). However, these methods only optimize intra-batch length variance; they do not disaggregate prefills versus re-prefills nor address the compute-memory interference we identify.

\section{Length-Aware-Prefill Serving (LAPS)}
%单卡disaggregate，多卡disaggregate

\begin{figure*}
    \centering
    \includegraphics[width=\linewidth]{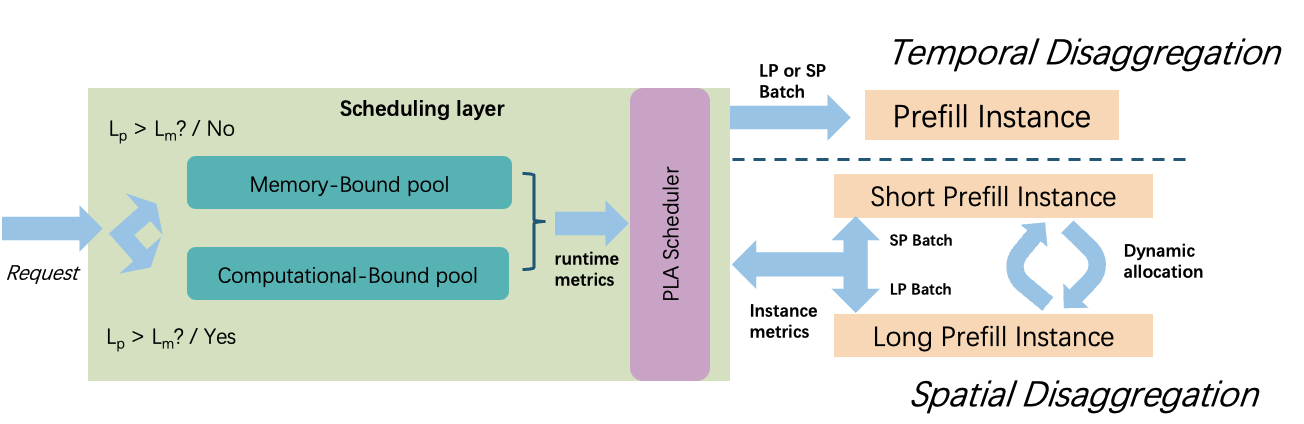}
    \caption{Resource utilization during multi-turn LLM inference.
Long-context requests saturate tensor cores during prefill (compute-bound), while short, frequent requests and re-prefill stages are memory-bound with high HBM usage—illustrating the interference between compute- and memory-bound workloads in shared serving systems.}
    \label{fig:system_overview}
\end{figure*}

We develop LAPS to mitigate the interference between long-prefill and short-prefill requests in multi-turn LLM serving.
The interference stems from two major factors: (1) their heterogeneous computation characteristics, and (2) head-of-line blocking caused by unified batching.
Section 3.1 introduces the strategies we design to optimize high-concurrency short-prefill workloads, while Section 3.2 shows our queue- and instance-level disaggregation mechanism that isolates these request types to reduce interference. LAPS is built upon the prefill instance in the PD disaggregation architecture and extends it with a finer-grained disaggregation design within the prefill stage.

\subsection{Short Prefill Optimization}
During auto-regressive serving, the majority of end-to-end latency typically arises from the \emph{decode} stage. Consequently, most existing optimization efforts (e.g., PD disaggregation \cite{zhong2024distservedisaggregatingprefilldecoding}, CUDA Graph acceleration \cite{harish2007accelerating}, and router-based load balancing across decoding instances \cite{hu2024routerbenchbenchmarkmultillmrouting,jain2025intelligentrouterllmworkloads,stripelis2024tensoropera,jitkrittum2025universal}) have been designed for the decode phase. However, as the diversity of LLM workloads grows (e.g., agent decision-making, chain-of-thought reasoning, and multi-turn task planning), the prefill stage has become an increasingly significant bottleneck, yet its optimization potential remains largely overlooked. Despite this trend, optimization for short and multi-turn prefills remains unexplored.

In the \emph{decode} phase, serving systems widely adopt CUDA Graphs because token-by-token computation is highly repetitive. Each decoding step runs nearly identical kernels with stable batch shapes, while frequent small launches make CPU dispatch overhead non-negligible. As decoding adds one token per step and keeps a fixed graph structure, CUDA Graphs effectively eliminate CPU overhead and reduce latency, improving throughput and responsiveness. In contrast, the \emph{prefill} stage performs full-sequence embedding and attention, where input lengths and batch compositions vary greatly across requests. These dynamics make tensor shapes unstable, preventing CUDA Graph reuse. Prefill is also dominated by large attention GEMMs, making graph capture expensive and rarely amortized. Hence, mainstream serving systems avoid CUDA Graphs in prefill and instead rely on conventional kernel launches or fused-kernel optimizations.

In multi-turn dialogues, each user message triggers a \emph{re-prefill} step that encodes new tokens on top of the cached KV states. Unlike the initial long-prompt prefill, re-prefill excludes the system prompt and contains only new user inputs, resulting in shorter and more uniform sequences. This stable shape pattern matches CUDA Graph’s fixed-structure requirement. In practice, most re-prefill segments have only a few dozen tokens, allowing high graph reuse through padding or bucketization (e.g., lengths 8, 16, 32, 64). Compared with the highly dynamic long-prefill, this ``short-prefill" regime incurs much lower graph-construction cost and delivers greater performance gains.

\smallskip\noindent
\textbf{Graph capture and bucketization.}
From the characteristics of intensive \emph{re-prefill} workloads, speculative decoding and token-level routing—although not conventional PD inference—generate numerous short re-prefill requests. Drawing inspiration from EAGLE-2’s speculative decoding optimization \cite{li2024eagle2fasterinferencelanguage, li2025eagle3scalinginferenceacceleration}, LAPS pre-defines a grid of \emph{power-of-two} prompt-length-batch-size buckets (e.g., $L\in\{8,16,32,64,128,256\}$ and $B\in\{1,2,4,8,16,32,64\}$) \cite{gao2023tbdb}.
At system initialization, a CUDA Graph is captured for each bucket under the assumption of fixed operator topology and variable memory addresses. During inference, each short-prefill request is padded to the nearest bucket and grouped with others sharing the same $(L,B)$ configuration, thereby maximizing graph reuse with negligible memory overhead.

\begin{figure}
    \centering
    \includegraphics[width=\linewidth]{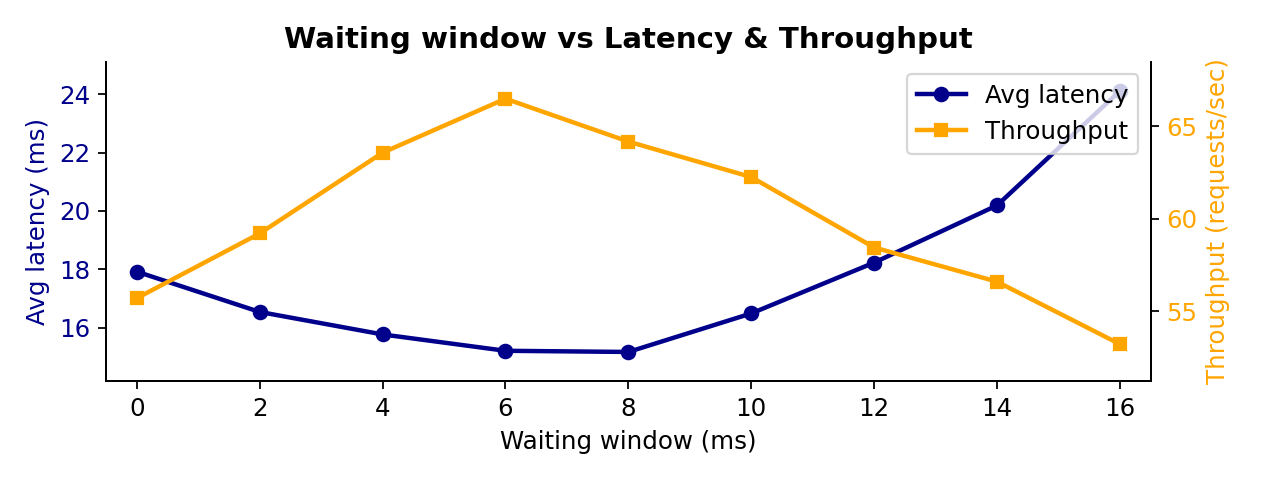}
    \caption{Average latency and throughput curves over varying waiting windows. The larger the waiting window, the more short-prefill requests will be batched. The serving system runs on an H200 GPU and a 14B model, with 64-way concurrency for short-prefill requests (prompt length less than 256 tokens).
    % We extend the serving system with configurable waiting windows to allow short-prefill requests to form larger batches. The figure illustrates how, on an H200 GPU running a 14B model, the average latency and throughput vary under 64-way concurrency for short requests (prompt length less than 256 tokens).
    }
    \label{fig:waiting_window}
\end{figure}

\smallskip\noindent
\textbf{Graph-aware memory-based batching.}
To maximize the benefit of CUDA Graphs for short-prefill workloads, LAPS optimizes batching with two goals: (i) reducing Graph launch frequency and (ii) increasing the reuse rate of large-batch Graphs. These are achieved through modestly extended waiting windows and graph fusion. Under high concurrency, slightly delaying batch formation allows more short-prefill requests to accumulate, improving overall efficiency when the saved launch overhead outweighs the waiting cost. Figure \ref{fig:waiting_window}  shows the latency-throughput trade-off under different window settings.

While current serving systems typically adopt a memory-constrained batching policy (i.e., aggregating requests until total tokens reach the GPU memory limit), LAPS enhances this approach with \emph{graph awareness}. During short-prefill batching, requests are grouped under the memory budget and aligned to the nearest captured Graph shape. 

%Our algorithm, Adaptive Wait-Depth (AWD), presented in Algorithm \ref{alg:awd_lite}, maintains two thresholds: a waiting window $W$ (maximum time to wait before dispatch) and a target depth $D$ (desired batch size aligned to a captured CUDA Graph shape). In each round, while $\mathrm{elapsed}<W$ and $\mathrm{depth}(B)<D$, it greedily packs short-prefill requests with a bucket-first rule (add requests that increase token count and fit memory with low padding); if the smallest batch slack $\min_{i\in B\cup\{\text{next}\}}\{\mathrm{DDL}_i-(t+\widehat S)\}\le\sigma$, it dispatches immediately to satisfy the SLA. Before dispatch, the batch is upgraded by $\textsc{NearestGraph}(B,\mathcal H,M)$ to the closest captured graph shape $G\!\in\!\mathcal H$ with $\mathrm{depth}(G)\!\ge\!\mathrm{depth}(B)$ and $\mathrm{mem}(G)\!\le\!M$; otherwise it falls back to the standard prefill kernel. After dispatch, if the batch fills the target ($d\!\ge\!D$) set $W\leftarrow\mathrm{clip}(\tau,[W_{\min},W_{\max}])$ (use the observed fill time for the next round); else set $D\leftarrow d$. \emph{Terms:} $\mathrm{DDL}_i$ = absolute deadline of request $i$; $\mathrm{slack}_i=\mathrm{DDL}_i-t-\widehat S$ (remaining time after one dispatch step, with $\widehat S$ the service-time estimate); $\mathrm{depth}(B)$ = number of sequences in batch $B$; $\mathcal H$ = set of captured CUDA Graph shapes with known depth and memory; $\sigma$ = SLA trigger; $\mathrm{clip}(\cdot)$ bounds a value to $[W_{\min},W_{\max}]$.

Our Adaptive Wait-Depth (AWD) scheduler, shown in Algorithm~\ref{alg:awd_lite}, maintains two adaptive thresholds: a waiting window $W$ (the maximum time to wait before dispatch)  and a target depth $D$ (the desired batch size aligned to a captured CUDA Graph shape). 
During each scheduling round, AWD accumulates short-prefill requests until either the waiting window $W$ expires or the target depth $D$ is reached. 
Requests are greedily grouped by input length to minimize padding, and dispatched early if any request  is close to violating its deadline. 
Before dispatch, the batch is matched to the nearest captured CUDA Graph configuration to maximize graph reuse;  otherwise, the standard prefill kernel is used. After each dispatch, $W$ and $D$ are dynamically updated based on the observed fill time and actual batch size 
for the next scheduling round.

\begin{algorithm}[t]\small
\caption{AWD: Adaptive-Wait-Depth Batching (Short-Prefill)}
\label{alg:awd_lite}
\Input{captured shapes $\mathcal H$ (depth, mem), budget $M$, slack threshold $\sigma$, bounds $[W_{\min},W_{\max}]$, service est. $\widehat S$}

$W \leftarrow \clip\!\big(\min_i(\mathrm{DDL}_i - t - \widehat S),[W_{\min},W_{\max}]\big)$\;
$D \leftarrow \max_{G\in\mathcal H:\ \mem(G)\le M}\depth(G)$\;

\While{server running}{
  start timer;\; $B \leftarrow \varnothing$\;
  \While{$\mathrm{elapsed} < W$ \textbf{and} $\depth(B) < D$}{
    \If{$\min_{i\in B\cup\{\text{next}\}}(\mathrm{DDL}_i-(t+\widehat S)) \le \sigma$}{\Break\tcp{SLA}}
    add next short request (bucket-first; fit mem)\;
  }
  $G^\star \leftarrow \NearestGraph(B,\mathcal H,M)$\; \tcp{nearest captured shape}
  \eIf{$G^\star$ exists}{pad $B$ to $G^\star$}{use standard prefill kernel}
  dispatch $B$;\; $d \leftarrow \depth(B)$;\; $\tau \leftarrow$ time to reach $d$\;
  \eIf{$d \ge D$}{$W \leftarrow \clip(\tau,[W_{\min},W_{\max}])$}{$D \leftarrow d$}
}
\end{algorithm}

\subsection{Long-Short Prefill Disaggregation}
To fundamentally eliminate the interference between long- and short-prefill requests discussed above, we adopt a design philosophy inspired by PD disaggregation, which further disaggregates long-prefill (LP) and short-prefill (SP) requests. However, unlike PD disaggregation, where the prefill and decode stages exhibit strong temporal dependencies and KV-cache transfers, the two types of tasks are merely mutually exclusive in our LP/SP disaggregation, resulting in fewer constraints in the scheduling objectives. Consequently, practical PD instance scheduling must account for the capacity coordination between the prefill and decode cluster, as well as the effective interconnect bandwidth when designing resource allocation strategies, but our design provides a larger design space for scheduling strategies that can adapt to the physical compute resources and workload characteristics. To this end, LAPS implements two complementary schedulers: a temporal disaggregation scheduler for single-instance prefill execution, and a spatial disaggregation scheduler for multi-instance prefill coordination. Figure \ref{fig:system_overview} presents the system overview.

Disaggregating prefill execution eliminates direct interference between compute-bound long-prefill and memory-bound short-prefill tasks; however, static separation alone cannot accommodate dynamic workload variations. In real-world deployments, the ratio of long to short requests fluctuates over time, and requests within each category exhibit heterogeneous lengths and deadlines. To address this, LAPS introduces a hierarchical scheduling layer: a temporal disaggregation scheduler is employed within each single prefill instance to manage intra-instance prioritization, while a spatial disaggregation scheduler operates across multiple prefill instances to coordinate inter-instance resource allocation.

It is worth noting that the disaggregation design further amplifies the benefits of CUDA Graphs for short-prefill workloads.
In mixed long/short prefill instances, a unified queue containing both request types limits the ability of short-prefill requests to form large, graph-aligned batches. In contrast, by maintaining two independent queues under the disaggregated design, LAPS can determine at request arrival whether CUDA Graph execution should be applied, thereby reducing batching delay and minimizing shape heterogeneity. In mixed queues, the large length disparity between long and short requests leads to excessive padding, lowering the Graph shape hit rate and GPU memory efficiency. After LP/SP separation, requests within each instance exhibit a more concentrated length distribution, improving both Graph reuse and throughput. As a result, the system can achieve higher CUDA Graph reuse and significantly reduce padding overhead.

\smallskip
\noindent
\textbf{Mutual exclusion.}
Prefill execution is \emph{disaggregated by length} at the instance level: each instance exclusively executes one type of prefill task, either short prefill (memory-bound) or long prefill (compute-bound). This mutual exclusion ensures that within an instance, GPU resources are never shared between the two classes, completely avoiding interference arising from scheduling strategies and heterogeneous computational characteristics. All requests are first classified by prompt length $L_p$ using the boundary point $L_m$. Each class maintains an independent queue: a short queue $Q_s$ and a long queue $Q_l$.

In real-world scenarios, inference tasks can be categorized into two major types depending on whether an individual request carries a strict TTFT requirement:

\textit{(a) SLA-constrained mode:}
Each request $i$ has an absolute deadline $\mathrm{DDL}_i$, and the scheduler jointly considers
\emph{SLA urgency} and \emph{CUDA Graph efficiency}.
At the beginning of each decision epoch $t$, we compute two candidate waiting windows and choose the tighter one:
\[
W(t)=\clip\!\Big(\min\{W_{\mathrm{SLA}}(t),\,W_{\mathrm{GR}}(t)\},\,[W_{\min},W_{\max}]\Big).
\]
The SLA window 
\[
W_{\mathrm{SLA}}(t)=\max\!\Big\{0,\,\min_{i\in\mathcal{Q}_s(t)}(\mathrm{DDL}_i-t-\widehat S)-\delta\Big\}
\]
represents the last safe time to wait before any pending short-prefill request would violate its deadline after one prefill step of duration~$\widehat S$ (with a small safety margin~$\delta$).
The Graph window 
\[
W_{\mathrm{GR}}(t)\approx\frac{\max\{0,\,D-\depth(B)\}}{\max\{\hat r_s,\,\varepsilon\}}
\]
is the expected time to reach the target batch depth~$D$ aligned to the nearest captured CUDA Graph shape, under the estimated short-request arrival rate~$\hat r_s$.
During batching, if the smallest batch slack $\min_{i\in B\cup\{\text{next}\}}(\mathrm{DDL}_i-(t+\widehat S))\le\sigma$ or the head-of-line wait exceeds~$T_{\max}$, we dispatch immediately.
Thus, SLA pressure shortens the waiting window when deadlines are tight, whereas under low SLA pressure, the scheduler may wait up to~$W_{\mathrm{GR}}$ to aggregate a larger batch and improve CUDA Graph reuse.
Long-prefill dispatch continues to advance a single request by fixed-size chunks~$C_l$, and each instance remains exclusive to either short or long mode.

\textit{(b) Deadline-free mode:}
For offline tasks like dataset distillation \cite{lei2023comprehensive}, each request does not have a preset deadline, and the policy reduces to \emph{token-max} under the same feasibility constraints.
The scheduler forms large, shape-similar short-prefill batches to fill the nearest captured CUDA Graph bucket (admit when $\mathrm{tok}(B)\ge M_s$), while long-prefill dispatches a single request with large fixed-size chunks $C_l$ to sustain high arithmetic intensity and maximize throughput.

\smallskip
\noindent\textbf{Temporal disaggregation mode for single instance.}
LAPS adopts a temporal disaggregation mode, where each GPU instance is dedicated exclusively to either short- or long-prefill execution.
Two global queues $Q_s$ and $Q_l$ are maintained for short and long requests, respectively, and each instance pulls tasks only from its own queue. Scheduling decisions within each instance follow the policies described in the previous section: \emph{SLA-first} (near-deadline priority) when deadlines are active, and \emph{token-max} (CUDA Graph aggregation) when no deadline is preset. This exclusive-per-class execution avoids long-short interference and ensures stable prefill latency under both modes.

\smallskip
\noindent\textbf{Spatial disaggregation mode for multiple instances.}
In the multi-instance setting, LAPS employs a controller to dynamically balance short- and long-prefill workloads across $N$ GPU instances (see Algorithm \ref{alg:inst-pressure}).
Two independent instance pools are maintained: $n_s$ short-prefill instances and $n_l=N-n_s$ long-prefill instances.
At each control interval, the controller monitors the queue backlog, SLA deviation, and GPU utilization of each instance to estimate its load pressure.
It then compares the overall pressures of the two pools and, after a cool-down period, migrates at most one instance between them when the imbalance exceeds a threshold.
This simple feedback control stabilizes P99 latency, prevents oscillation, and keeps GPU utilization high with negligible overhead.

\begin{algorithm}[t]
\caption{Lightweight Instance-Pressure Controller}
\label{alg:inst-pressure}
\SetKwInOut{Input}{Input}
\Input{Total $N$; current $(n_s,n_l)$; control period $\Delta t$; cool-down $T_{\mathrm{cool}}$; hysteresis $\tau$; min allocation $n^{\min}$; weights $(\alpha,\beta,\gamma)$; robust aggregator $A(\cdot)$}
$t_{\mathrm{last}}\leftarrow -\infty$\;
\While{server running}{
  \textbf{sleep} $\Delta t$\;
  \tcp{collect per-instance signals for both pools}
  \ForEach{instance $k$ in SHORT pool}{
    measure $q_k, e_k, u_k$;\quad $\psi_k \leftarrow \alpha\,q_k+\beta\,e_k-\gamma\,u_k$;
  }
  \ForEach{instance $k$ in LONG pool}{
    measure $q_k, e_k, u_k$;\quad $\psi_k \leftarrow \alpha\,q_k+\beta\,e_k-\gamma\,u_k$;
  }
  \tcp{robust pool pressures ($P90$)}
  $P_s \leftarrow A(\{\psi_k: k\in \text{SHORT}\})$\;
  $P_l \leftarrow A(\{\psi_k: k\in \text{LONG}\})$\;
  \If{$\mathrm{now}-t_{\mathrm{last}}<T_{\mathrm{cool}}$}{\textbf{continue}}\;
  \tcp{single-step hill-climb with hysteresis and safeguards}
  \uIf{$P_s > (1+\tau)\,P_l$ \textbf{ and } $n_l>n^{\min}$}{
     migrate one instance: $n_s\!\leftarrow\!n_s+1$;\ $n_l\!\leftarrow\!n_l-1$;\ $t_{\mathrm{last}}\!\leftarrow\!\mathrm{now}$;
  }
  \uElseIf{$P_l > (1+\tau)\,P_s$ \textbf{ and } $n_s>n^{\min}$}{
     migrate one instance: $n_l\!\leftarrow\!n_l+1$;\ $n_s\!\leftarrow\!n_s-1$;\ $t_{\mathrm{last}}\!\leftarrow\!\mathrm{now}$;
  }
}
\end{algorithm}
%need to plot a diagram to illustrate the system workflow

\section{Experiments}
In both online LLM serving and offline LLM tasks (e.g., dataset distillation), the system must handle highly concurrent requests with heterogeneous task types. 

We implement and deploy LAPS on an NVIDIA H200 GPU as well as on an 8$\times$H200 multi-GPU cluster. 
We build it upon SGLang by extending $\sim$2K lines of code.
We evaluate the prototype system under both single- and multi-GPU settings using a variety of workload patterns:
\begin{enumerate}
    \item Online task: High-concurrency multi-turn conversations with long/short prompts;
    \item Offline task: Full dataset distillation without deadline constraints on single requests.
\end{enumerate}

Our evaluations focus on multi-turn conversational workloads. We use LMsys-Chat-1M \cite{zheng2024lmsyschat1mlargescalerealworldllm} and ShareGPT \cite{zheng2023judging} as our datasets,
which consist of large-scale, real-world human-assistant conversations collected from ChatGPT and LMsys platforms.

\paragraph{Metrics.}
We collect several key metrics for the prefill stage, including TTFT, P90 latency, average request per second (RPS), and SLO violation rate \cite{wang2024revisiting}.

\begin{figure*}[ht!]
    \centering
    % 上图（temporal disaggregation, instance = 1）
    \includegraphics[width=0.9\linewidth]{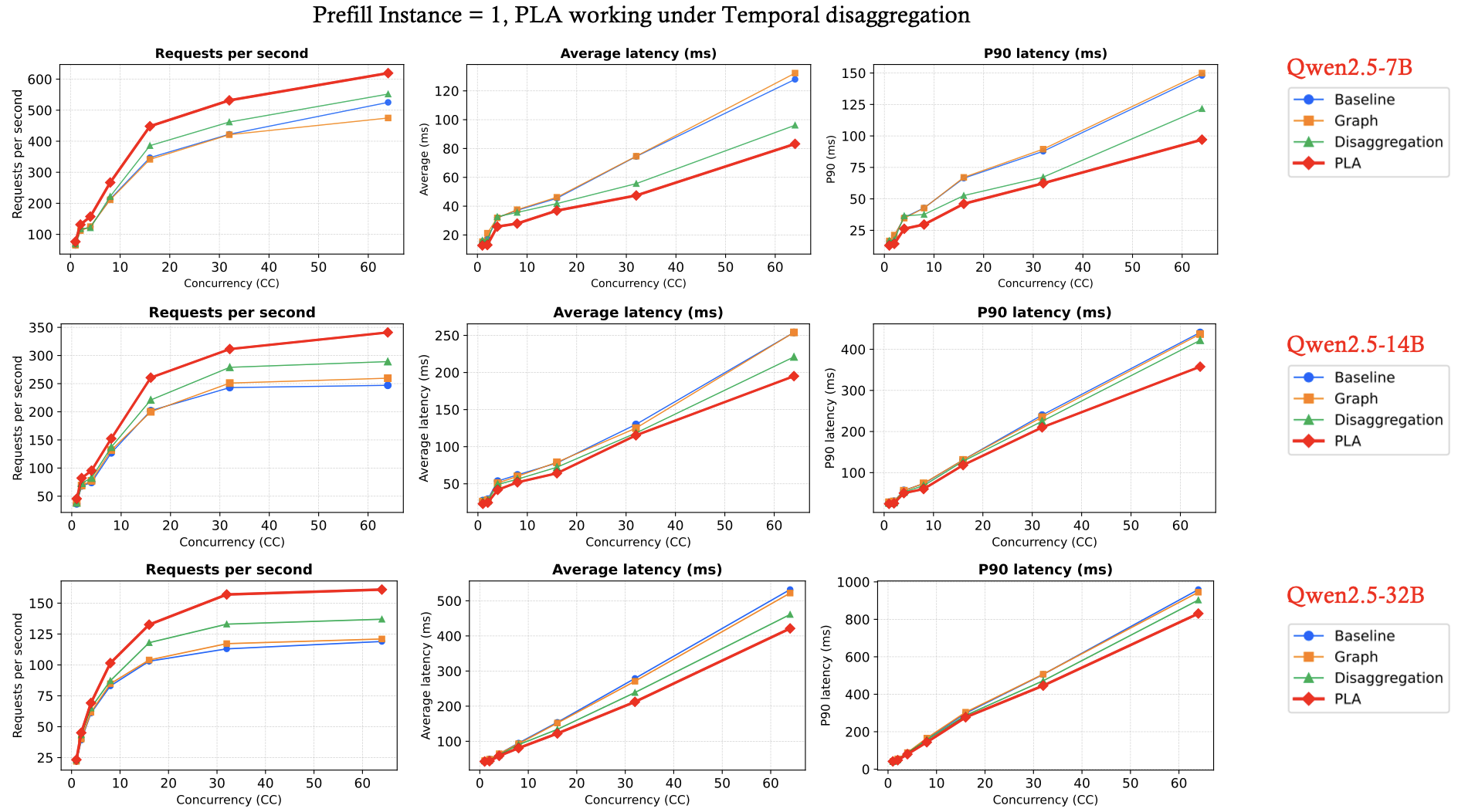}\\[0.5em]
    % 下图（spatial disaggregation, instance = 8）
    \includegraphics[width=0.9\linewidth]{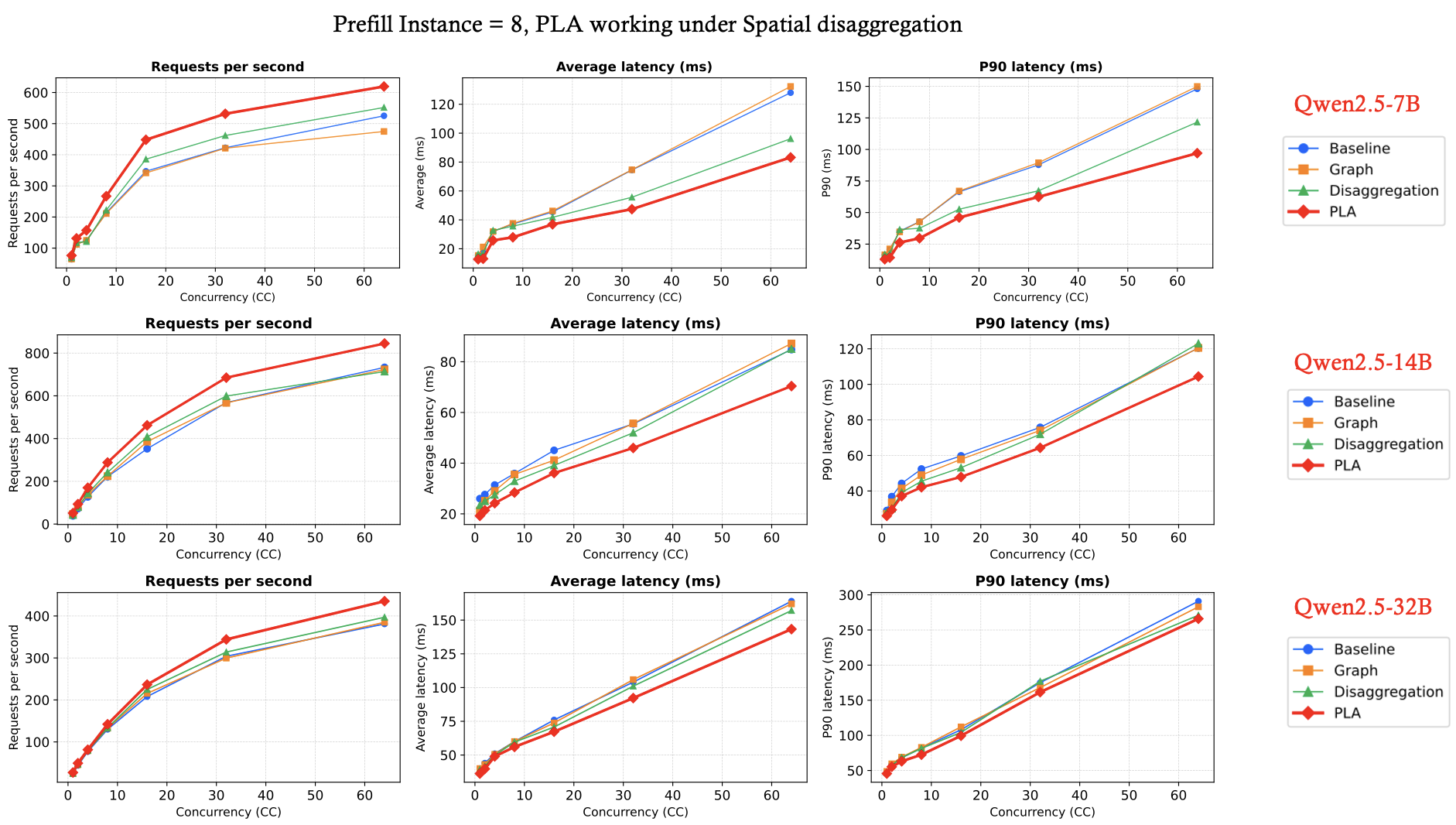}
    \caption{
        Comparison of LAPS and SGLang on one H200 node. 
        The top two lines of figures correspond to the \textit{temporal disaggregation} setting 
        (with prefill instance of 1), 
        while the bottom two lines of figures correspond to the \textit{spatial disaggregation} setting 
        (with prefill instance of 8). 
        We report RPS, average latency, and P90 latency under four configurations:
        \textit{Vanilla SGLang PD disaggregation (blue line)}, 
        \textit{LAPS (only CUDA Graphs enabled, orange line)}, 
        \textit{LAPS (only disaggregation enabled, green line)}, 
        and \textit{Full LAPS (red line)}.
    }
    \label{fig:result_1}
\end{figure*}

\paragraph{Baselines.}
We compare LAPS against SGLang and vLLM, both are state-of-the-art LLM serving systems. 
\textit{SGLang} is a widely adopted serving system in both academia and industry; it implements continuous batching to improve throughput and radix-attention~\cite{zheng2024sglangefficientexecutionstructured} to mitigate memory fragmentation during KV-cache allocation.  However, neither SGLang nor vLLM supports CUDA Graph during the prefill phase, and their batching policies rely solely on available memory capacity, so they cannot adjust their batching strategies according to the workload characteristics.

%\subsection{multi-turn Short request flush}

%We first use the ShareGPT dataset, selecting multi-turn dialogue samples whose individual requests contain no more than 256 tokens. These requests are processed respectively with SGLang, vLLM, and LAPS.  We evaluate each system under concurrency levels ranging from 1 to 64. As shown in Figure~\ref{fig:short-prefill}, under the \emph{short prefill} setting (average request length $<$ 256 tokens), LAPS achieves an average throughput improvement of 20.5\,\% over SGLang and 23.7\,\% over vLLM. At the same time, the latency percentiles (P50, P90, P99) are reduced by 23\,\%, 19\,\%, and 17\,\%, respectively. These results demonstrate that, for short-prefill workloads, leveraging CUDA Graphs can significantly reduce the kernel-launch overhead incurred by repeated attention computations during each prefill operation.

%only short的并发在不同TP下的latency和吞吐报告
\subsection{Numerical Results and Analysis}
%真实dataset处理完成时间，3种model，长短分开发，一起发，LAPS server发，SGlang + Vllm

Figure \ref{fig:result_1} compares the performance of LAPS with SGLang (with PD disaggregation) and its two partial variants (LAPS with CUDA Graph only and LAPS with Disaggregation only) under a sustained client load with varying concurrency levels (from 1 to 64). The requests are drawn from real multi-turn conversations in the ShareGPT-4 dataset.
We evaluate three models, Qwen2.5-7/14/32B, under both the single-instance (temporal disaggregation) and 8-instance (spatial disaggregation) settings. 

LAPS consistently outperforms SGLang and its two partial variants across all key metrics: RPS, average latency, and P90 latency. 
The benefits of LAPS's scheduling mechanism and CUDA Graph optimization become more pronounced under high concurrency.  Specifically, LAPS achieves up to 20\% and 33\% higher RPS than the baseline in the single-prefill instance and 8-prefill instance settings, while reducing average latency by 20\% respectively.

It is worth noting that, in some configurations, enabling CUDA Graphs alone yields limited improvements and can even degrade throughput, 
as the overhead of graph eligibility checking and graph launching becomes non-negligible. 
Enabling disaggregation, however, allows the system to dynamically adjust the waiting window size and form larger batches, 
thereby amplifying the effective performance gains from CUDA Graph execution and making its scheduling and launch overhead negligible.

\begin{figure}[h!]
    \centering
    \includegraphics[width=0.85\linewidth]{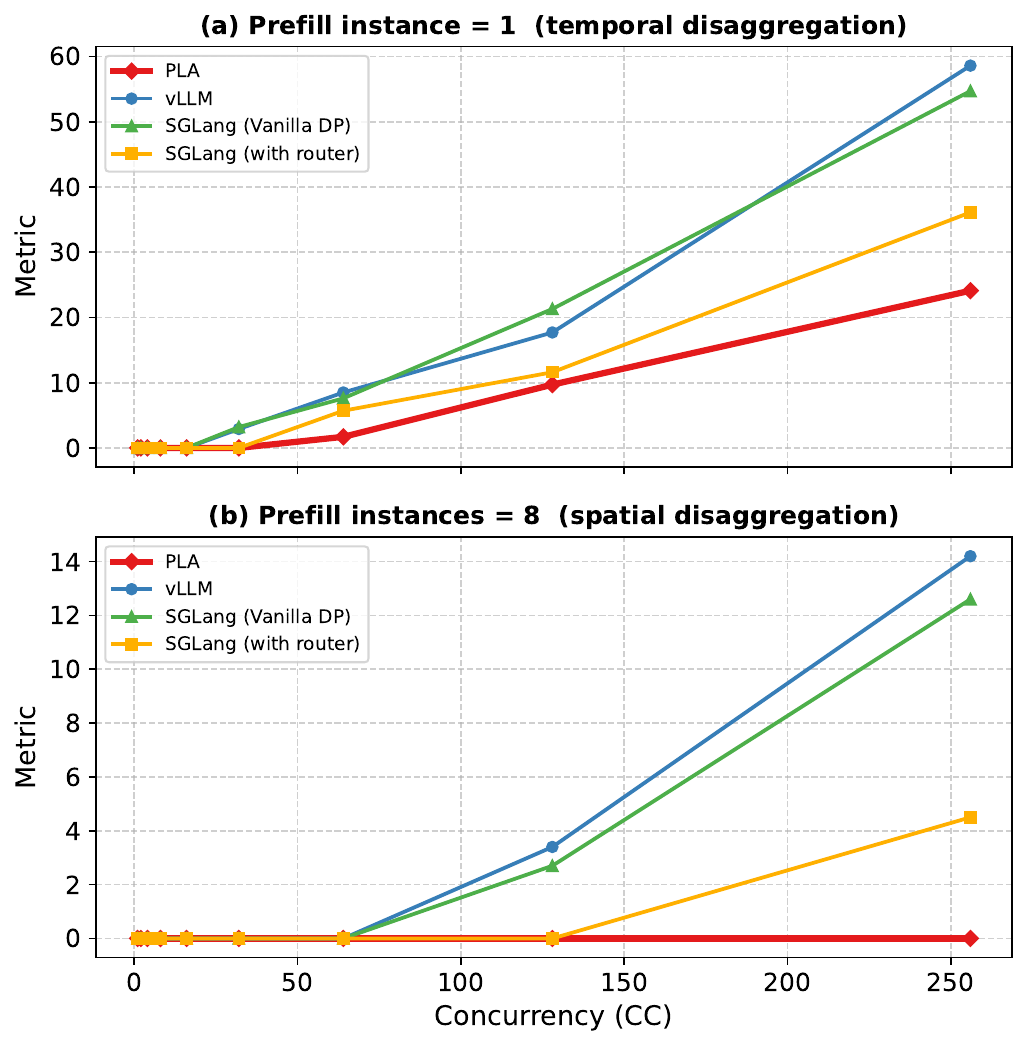}
    \caption{
SLO violation rate under varying client concurrency levels using the LMsys-Chat-1M dataset. Results are shown for \textit{LAPS}, \textit{SGLang (PD disaggregation)}, \textit{SGLang (PD disaggregation with router)}, and \textit{vLLM (PD disaggregation)} under two settings: 
(top) single-instance (temporal disaggregation) and (bottom) 8-instance (spatial disaggregation).
}
    \label{fig:SLO}
\end{figure}

In Figure~\ref{fig:SLO}, we use the LMsys-Chat-1M dataset and assume that request arrivals follow a Poisson process with an average arrival rate $\lambda$, while each request’s service time follows an empirical distribution measured from model execution.  We set the TTFT SLO to 0.4s and vary the client-side concurrency levels to observe the actual SLO violation rate. \textit{SGLang} supports data parallel (DP) serving based on a router that dispatches requests to different workers using either round-robin or load-balancing strategies; however, the router is unaware of the SLOs of individual requests.  As shown in the figure, within a single instance, \textit{LAPS} reduces the SLO violation rate by approximately 10\% compared with \textit{SGLang (PD disaggregation) with router}, and by around 30\% compared with \textit{Vanilla DP}. 
Under the 8-instance spatial disaggregation setting, \textit{LAPS} achieves zero SLO violations, whereas \textit{SGLang} with router still exhibits a 4.7\% violation rate.

\begin{figure}
    \centering
    \includegraphics[width=0.9\linewidth]{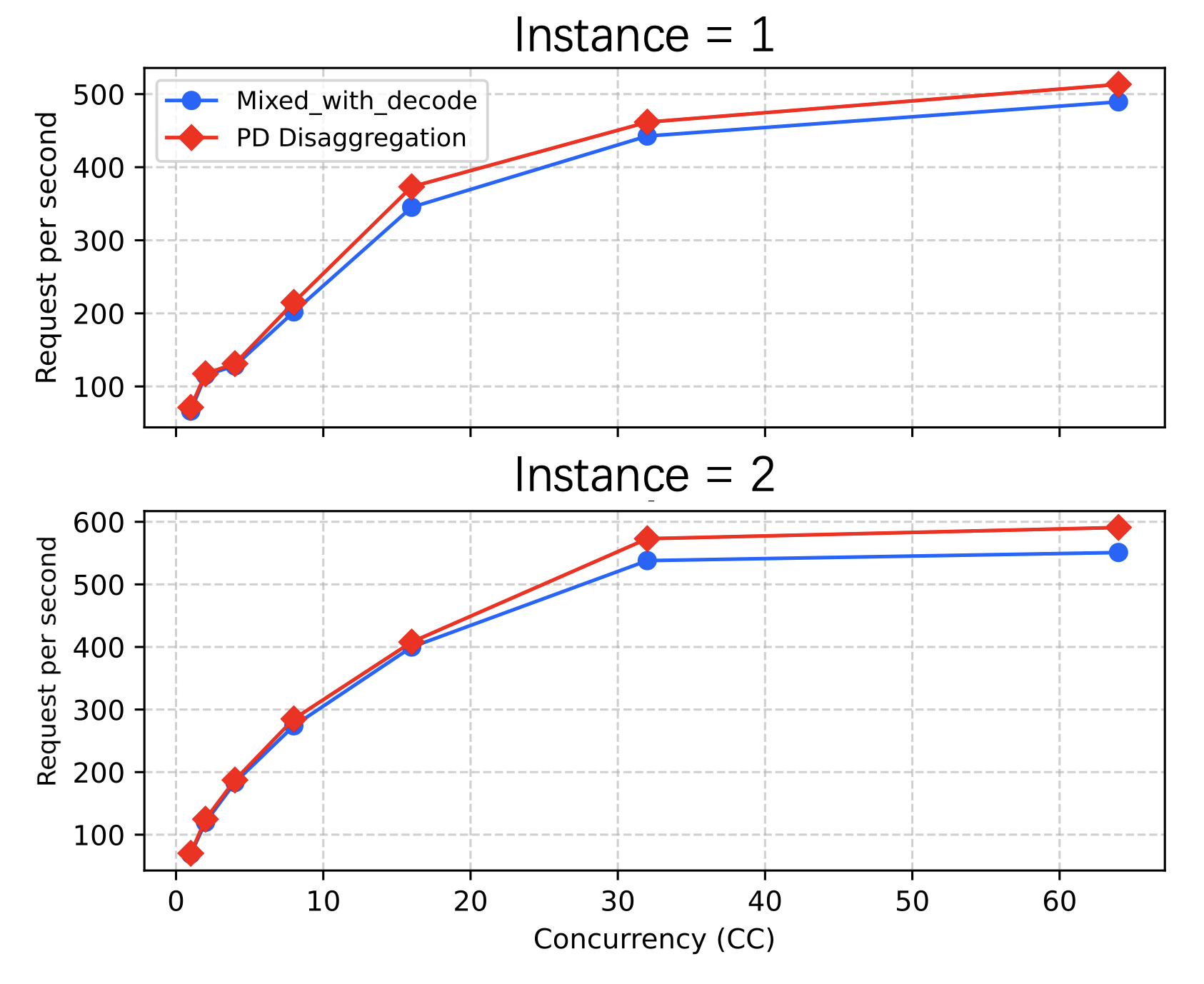}
    \caption{
Comparison of prefill throughput between PD disaggregation and Mixed with Decode across different concurrency levels under single- and 2-instance settings.
}
    \label{fig:decode_interfere}
\end{figure}

In Figure~\ref{fig:decode_interfere}, we evaluate the compatibility of LAPS under non-PD-disaggregated settings by mixing prefill and decode requests at different concurrency levels. 
In both single-instance and multi-instance configurations, the request-per-second (RPS) of prefill requests decreases under the \textit{Mix with Decode} condition, indicating that LAPS can fully exploit the throughput benefits of CUDA Graphs for short prefill requests only within the PD-disaggregated architecture. 
When mixed with decode workloads, the lack of continuous batching introduces additional inter-batch latency, resulting in degraded overall performance.

As shown in Table~\ref{tab:decode_speed}, we evaluate the distillation task where 
each request has no strict deadline (i.e., the waiting window can be relatively large). 
Under this setting, with four prefill instances and a decoding length of 1K tokens, 
LAPS achieves about an 8\% reduction in time consumption compared to SGLang (Vanilla PD disaggregation).
\vspace{-0.2cm}

\begin{table}[t]
\centering
\caption {End-to-end time of PD-disaggregated serving with 4 prefill and 4 decode instances, distilled on two dialogue datasets.}
\label{tab:decode_speed}
\renewcommand{\arraystretch}{1.05}
\setlength{\tabcolsep}{4pt}
\vspace{2mm}
\begin{tabular}{p{0.48\linewidth}cc}
\toprule
\textbf{System Setting} & \textbf{LMSys} & \textbf{ShareGPT} \\
\midrule
4 P (SGLang) + 4 D & 18{,}963s & 25{,}132s \\
4 P (LAPS) + 4 D & 17{,}572s & 23{,}041s \\
\textit{Improvement} & \textit{7.3\%} & \textit{8.3\%} \\
\bottomrule
\end{tabular}
\end{table}

\vspace{-0.1cm}

\subsection{Cost analysis}

In LAPS deployment, CUDA Graphs are captured into memory during initialization.
Each graph is bound to a fixed kernel configuration and cannot adapt to dynamic kernel sizes, so multiple graphs must be captured to cover different token lengths and batch sizes.
Each prefill step introduces lookup and selection overhead, and thus, the number of graphs must be limited to balance memory usage and performance.
We measure single-graph sizes of 228 MB, 240 MB, and 277 MB for the 7B, 14B, and 32B models,
showing that graph size is largely insensitive to model scale. When the system is initialized for the first time, it needs to capture kernels and the KV-cache operations layer by layer, which introduces a certain startup overhead. Experiments show that capturing a single prefill graph incurs an initialization overhead of approximately 8-12 seconds.

%内存开销，启动开销

\section{Conclusion}
In this paper, we propose \textit{LAPS}, a prefill-length-aware LLM serving system built on the PD disaggregation paradigm to optimize heterogeneous multi-turn conversational workloads.
By separating long- and short-prefill requests, LAPS eliminates compute–memory interference in the prefill stage.
Its adaptive scheduler (AWD) and CUDA Graph–based execution improve batching efficiency and reduce short-prefill latency.
Supporting both temporal and spatial disaggregation, LAPS scales across single- and multi-prefill-instance deployments.
Experiments on real-world datasets show that LAPS achieves higher throughput and lower latency than state-of-the-art frameworks (e.g., SGLang and vLLM under PD disaggregation), demonstrating its effectiveness under high concurrency.

\newpage

\bibliography{example_paper}
\bibliographystyle{mlsys2025}

%%%%%%%%%%%%%%%%%%%%%%%%%%%%%%%%%%%%%%%%%%%%%%%%%%%%%%%%%%%%%%%%%%%%%%%%%%%%%%%
%%%%%%%%%%%%%%%%%%%%%%%%%%%%%%%%%%%%%%%%%%%%%%%%%%%%%%%%%%%%%%%%%%%%%%%%%%%%%%%
% SUPPLEMENTAL CONTENT AS APPENDIX AFTER REFERENCES
%%%%%%%%%%%%%%%%%%%%%%%%%%%%%%%%%%%%%%%%%%%%%%%%%%%%%%%%%%%%%%%%%%%%%%%%%%%%%%%
%%%%%%%%%%%%%%%%%%%%%%%%%%%%%%%%%%%%%%%%%%%%%%%%%%%%%%%%%%%%%%%%%%%%%%%%%%%%%%%

%%%%%%%%%%%%%%%%%%%%%%%%%%%%%%%%%%%%%%%%%%%%%%%%%%%%%%%%%%%%%%%%%%%%%%%%%%%%%%%
%%%%%%%%%%%%%%%%%%%%%%%%%%%%%%%%%%%%%%%%%%%%%%%%%%%%%%%%%%%%%%%%%%%%%%%%%%%%%%%

\end{document}